\documentclass{extarticle}

\usepackage{times,pslatex}
\usepackage{algorithm,algorithmic,float}
\usepackage{mhchem,babel}
\usepackage{amsfonts, amsmath, amssymb, amsthm}
\usepackage{booktabs}
\theoremstyle{definition}

\newtheorem{definition}{Definition}

\newtheorem{example}{Example}
\theoremstyle{plain}

\usepackage{enumitem}
\usepackage{natbib}
\usepackage{url}
\usepackage[dvipsnames]{xcolor}
\usepackage[margin=3cm,letterpaper]{geometry}
\usepackage[colorlinks=true,linkcolor=RoyalBlue,
citecolor=blue,
urlcolor=blue]{hyperref}

\usepackage[multiple]{footmisc}

\DeclareMathOperator{\obs}{obs}

\usepackage{graphicx}
\graphicspath{{images/}}

\newcommand{\biomodel}[1]{\textsl{#1}}
\newcommand{\cl}{constrained linear lumping}

\title{CLUE: Exact maximal reduction of kinetic models \\by constrained lumping of differential equations}

 \author{Alexey Ovchinnikov\footnote{\url{aovchinnikov@qc.cuny.edu}, Department of Mathematics, CUNY Queens College, Queens, NY, 11367 and CUNY Graduate Center, New York, NY, 10016, USA}, Isabel  P\'erez Verona\footnote{\url{isabel.perez@alumni.imtlucca.it}, IMT School for Advanced Studies, Lucca, 55100, Italy},
 Gleb Pogudin\footnote{\url{gleb.pogudin@polytechnique.edu}, LIX, CNRS, \'Ecole Polytechnique, Institute Polytechnique de Paris, Palaiseau, 91120, France} \footnote{Corresponding author. The authors are ordered alphabetically.}, Mirco Tribastone\footnote{\url{mirco.tribastone@imtlucca.it}, IMT School for Advanced Studies, Lucca, 55100, Italy}}
 
\begin{document}
 \date{}
 \maketitle

\begin{abstract}\noindent\textbf{Motivation:} Detailed mechanistic models of biological processes can pose significant challenges for analysis and parameter estimations due to the large number of equations used to track the dynamics of all distinct configurations in which each involved biochemical species can be found. Model reduction can help tame such complexity by providing a lower-dimensional model in which each macro-variable can be directly related to the original variables.  \\
\textbf{Results:} We present CLUE, an algorithm for exact model reduction of systems of polynomial differential equations by constrained linear lumping. It computes the smallest dimensional reduction as a linear mapping of the state space such that the reduced model preserves the dynamics of user-specified linear combinations of the original variables. 
Even though CLUE
works with nonlinear differential equations, it is based on linear algebra tools, which makes it  applicable to high-dimensional models.
Using case studies from the literature, we show how
CLUE can  substantially lower model dimensionality and help extract biologically intelligible insights from the reduction.\\
\textbf{Availability:} An implementation of the algorithm and relevant resources to replicate the experiments herein reported are freely available for download at \href{https://github.com/pogudingleb/CLUE}{https://github.com/pogudingleb/CLUE}.\\
\textbf{Supplementary information:} Supplementary materials are enclosed below.
\end{abstract}

%%%%%%%%%%%%%%%%%%%%%%%%%%%%%%%%%%%%%%%%%%%%%%%%%%%%%%%%%%

\section{Introduction}

 Kinetic
models of biochemical systems hold the promise of being able to unravel mechanistic insights in living cells as well as predict the behavior of a biological process under unseen circumstances, which is a fundamental premise for many applications including control and synthesis. 

In order to obtain an accurate model, however, it is often necessary to incorporate a substantial amount of detail about the specific mechanisms of interaction between the different components of a biological system. In many cases, this may lead to an overall representation which hinders physical intelligibility. For example, a mechanistic description of protein  phosphorylation---a basic, ubiquitous process in signaling pathways~\citep{PAWSON2005286}---may yield models with a combinatorially large number of variables, particularly in the case of multisite phosphorylation~\citep{FEBS:FEBS7027}. 

Model reduction represents a promising class of methods designed for obtaining a lower-dimensional representation that retains some dynamical features of interest to the modeler. The substantial body of research available is motivated by the fact that it is a cross-cutting concern throughout many scientific and engineering disciplines to be able to effectively work with simple but accurate models of complex systems. Specifically, for applications to systems biology the availability of a smaller model can be particularly beneficial in order to reduce the number of kinetic parameters~\citep{doi:10.1111/j.1742-4658.2006.05485.x}, whose measurements and calibration is a well-known hindrance, see, e.g.,~\citep{doi:10.1098/rsif.2017.0237}.  

Techniques based on balanced truncation and singular value decomposition can dramatically lower the dimensionality of a model with small approximation errors~\citep{antoulas}. Since the reduced model preserves the input/output behavior, it can be conveniently used in place of the original model to speed up the computation time of a numerical simulation. However, the coordinate transformation typically destroys the structure, leading to a loss of physical interpretability of the model. This is
recognized as an important property to be maintained in applications to systems biology, especially if the model is used to validate mechanistic hypotheses~\citep{10.1093/bioinformatics/btn035,21899762,Apri201216}. 

For models of biochemical systems, many reduction methods are based on exploiting time-scale separation~\citep{okino1998}. One of the most 
well-known approaches is
quasi-steady-state approximation~\citep{SegelS89},
in which, roughly speaking, ``fast'' variables can be approximated as reaching their stationary values such that they can be replaced
by constants (solutions of the
associated system of equations) in the dynamical model for the ``slow'' variables. Another class of reduction techniques based on sensitivity analysis studies how model parameters and variables affect the desired output, suggesting the elimination of the least influential ones~\citep{Snowden:2017aa}.    

Exact model reduction aims at lowering  dimensionality without introducing approximation errors in the reduced model. Conservation analysis  detects linear combinations of variables that remain constant at all times~\citep{10.1093/bioinformatics/bti800}. Exact lumping is a more general approach whereby it is possible to write a self-consistent system of dynamical equations for a set of macro-variables in which each macro-variable represents a combination of the original ones~\citep{okino1998}. Linear lumping, known as early as in~\citet{Wei:1969aa}, expresses such combinations as a linear mapping on the original state variables. To maintain some degree of physical interpretability, the lumping may be restricted only to a part of the state space. \citet{LI199195} allow the specification of linear combination of variables that ought to be preserved. More recently, \citet{pnas17} presented a lumping algorithm that identifies a partition of the state variables such that in the lumped system each macro-variable represents the sum of the original variables of a block. Specialized lumping criteria have also been studied for classes of biochemical models for signaling pathways, e.g.,~\citep{Borisov2005951,16430778,Feret21042009}, for example by analyzing higher-level descriptions such as rule-based systems from which ordinary differential equation (ODE) models can be generated~\citep{Danos200469,Blinov22112004}. {However, these approaches identify specific types of exact linear lumping, namely if the macro-variables represent sums of the original variables of the original model (but not necessarily induced by a partition of the state space as in~\citet{pnas17}).}

{There are connections between exact model reduction and observability/identifiability (see~\citep{Miao2011, Villaverde2019} for discussion of these notions in the biological contexts).
If a model admits an exact reduction preserving the observed variables, then it is not observable because the states with the same image under the reduction are indistinguishable.
On the other hand, the reduction provides a reparametrization with {fewer degrees of freedom with respect to the observations}.
Moreover, since the reduction is exact, the state (or parameter) identification can be carried out for the reduced model without any loss of information but at lesser computational costs, especially for identification methods sensitive to the model dimension (e.g., sampling-based ones~\citep{MCMC}).
}

Here we present CLUE, an algorithm for constrained linear lumping, applicable to models as ODEs with polynomial derivatives. 
The constraints represent the linear combinations of state variables that ought to be maintained in the reduced model, similarly to~\citet{LI199195}. The algorithm hinges on the fundamental observation by the same authors~\citep{LiRabitz, LI199195} that exact lumpings correspond to the subspaces that are invariant under the Jacobian of the ODE system.
For finding these subspaces, \citet{LiRabitz,LI199195} suggest two ways: (i) produce a finite set of constant matrices such that every common invariant subspace of these matrices would be invariant for the Jacobian (but not necessarily vice versa); and (ii) find eigenvectors and eigenvalues of the Jacobian symbolically, and explore their combinations.
In the former approach, the obtained set of matrices might be too restrictive, so a lumping might not be found even if there is one.
The latter approach is limited to small sized systems because it involves finding symbolic expressions to the eigenvalues of a nonconstant matrix (the Jacobian of the system) and requires human intervention for exploring various combinations of these eigenvectors (for example, \cite[\S 4, Example~2]{LiRabitz}).

Our main contribution is twofold. First, we  provide a set of constant matrices
whose common invariant subspaces are exactly the invariant subspaces of the Jacobian.
This allows us to obtain a fully algorithmic method for finding a constrained lumping based purely on linear algebra. 
Second, we improve the algorithm 
by eliminating redundant computation from the invariant subspace generation and by using modular computation to avoid intermediate expression swell. 
This enables the analysis of models with several thousands of equations on commodity hardware. 
Together, our results allow us to study large-scale biochemical models, of which we present a number of case studies, showing the degree of lumpability achieved as well as the physical interpretation of the reduced system.

%%%%%%%%%%%%%%%%%%%%%%%%%%%%%%%%%%%%%%%%%%%%%%%%%%%%%%%%%%%

\section{Approach and method}

\begin{definition}[Lumping]\label{def:lumping}
Consider a system of ODEs with polynomial right-hand side in the form
  \begin{equation}\label{eq:main}
    \dot{\mathbf{x}} = \mathbf{f}(\mathbf{x}),
  \end{equation}
  where $\mathbf{x} = (x_1, \ldots, x_n)^T$, $\mathbf{f} = (f_1, \ldots, f_n)^T$, and $f_1, \ldots, f_n \in \mathbb{R}[\mathbf{x}]$.
  We say that a linear transformation $\mathbf{y} = L\mathbf{x}$ with $\mathbf{y} = (y_1, \ldots, y_m)^T$, $L \in \mathbb{R}^{m \times n}$, and $\operatorname{rank}L = m$ is \emph{a lumping of~\eqref{eq:main}} if there exist $\mathbf{g} = (g_1, \ldots, g_m)^T$ with $g_1, \ldots, g_m \in \mathbb{R}[\mathbf{y}]$ such that 
  \[
    \dot{\mathbf{y}} = \mathbf{g}(\mathbf{y})
  \]
  for every solution $\mathbf{x}$ of~\eqref{eq:main}. We say that $m$ is \emph{the dimension of the lumping}. The variables $\mathbf{y}$ in the reduced system are called \emph{macro-variables}. 
\end{definition}

\begin{example}\label{ex:lumping}
Consider the system
  \begin{equation}\label{eq:ex_running}
      \dot{x}_1  = x_2^2 + 4x_2x_3 + 4x_3^2, 
      \ \ 
    \ \   \dot{x}_2  =  4x_3 - 2x_1, 
     \ \  \dot{x}_3  = x_1 + x_2.
  \end{equation}
  We claim that the matrix
  \begin{equation}\label{eq:lumping_matrix}
  L = 
  \begin{pmatrix}
    1 & 0 & 0\\
    0 & 1 & 2
  \end{pmatrix}
  \end{equation}
  gives a lumping of~\eqref{eq:ex_running} of dimension two.
  Indeed,
  \[
  \begin{pmatrix}
     \dot{y}_1\\
     \dot{y}_2
  \end{pmatrix} =
  \begin{pmatrix}
    \dot{x}_1\\
    \dot{x}_2 + 2\dot{x}_3
  \end{pmatrix} = 
  \begin{pmatrix}
    (x_2 + 2x_3)^2\\
    2x_2 + 4x_3
  \end{pmatrix} =
  \begin{pmatrix}
    y_2^2\\
    2y_2
  \end{pmatrix},
  \]
  so we can take $g_1(y_1, y_2) = y_2^2$ and $g_2(y_1, y_2) = 2y_2$.
\end{example}

The lumping matrix of~\eqref{eq:lumping_matrix} turns out to exactly preserve the solution of  variable $x_1$. 
In general, one considers a vector $\mathbf{x}_{\obs}$ of combinations of the original variables that are to be recovered in the reduced system; that is, $\mathbf{x}_{\obs}$ is a vector of linearly independent forms in $\mathbf{x}$ such that  $\mathbf{x}_{\obs} = A\mathbf{x}$. 
Then we say that a lumping $\mathbf{y} = L\mathbf{x}$ is a \emph{constrained linear lumping}  
if each entry of $\mathbf{x}_{\obs}$ is a linear combination of the entries of $\mathbf{y}$. 

\begin{example}
Using the system~\eqref{eq:ex_running}, setting
  \[
  \mathbf{x}_{\obs} = A \mathbf{x}, \qquad \text{with} \qquad A = \begin{pmatrix} 
    1 & 0 & 0 \\
    1 & 1 & 2 \\
  \end{pmatrix},
  \]
  we see that~\eqref{eq:lumping_matrix} is a \cl\ because
  \[
  \mathbf{x}_{\obs} = \begin{pmatrix} 
    x_1 \\
    x_1 + x_2 + 2x_3
  \end{pmatrix} = 
  \begin{pmatrix}
  y_1\\
  y_1 + y_2
  \end{pmatrix}.
  \]
  Instead, setting $\mathbf{x}_{\obs} = (x_2)$ does not give a constrained lumping for $L$ because $(0, 1, 0)$ does not belong to the row space of $L$.
\end{example}

For a given vector $\mathbf{x}_{\obs}$, there may be more than one \cl. 
We define two lumpings $\mathbf{y}_1 = L_1\mathbf{x}$ and $\mathbf{y}_2 = L_2\mathbf{x}$ to be equivalent if there exists an invertible matrix $T$ such that $L_1 = TL_2$. 
It is possible to prove that, for every nonzero vector $\mathbf{x}_{\obs}$, there exists a unique (up to equivalence) lumping of the smallest possible dimension.

\begin{algorithm}
\caption{Simplified algorithm for finding a constrained lumping of the smallest possible dimension}\label{alg:simplified}
\begin{description}
\item[Input ] a system $\dot{\mathbf{x}} = \mathbf{f}(\mathbf{x})$ of $n$ ODEs with a polynomial right-hand side and an $s \times n$ matrix $A$ over $\mathbb{R}$ of rank $s > 0$;
\item[Output ] a matrix $L$ such that $\mathbf{y} := L\mathbf{x}$ is a constrained lumping with observables $A\mathbf{x}$ of smallest possible dimension.
\end{description}

\begin{enumerate}[label = \textbf{(Step~\arabic*)}, leftmargin=*, align=left, labelsep=2pt]
    \item Compute $J(\mathbf{x})$, the Jacobian matrix of $\mathbf{f}(\mathbf{x})$.
    \item Represent $J(\mathbf{x})$ as $J_1m_1 + \ldots + J_Nm_N$, where $m_1, \ldots, m_N$ are distinct monomials in $\mathbf{x}$, and $J_1, \ldots, J_N$ are nonzero matrices over $\mathbb{R}$.
    \item\label{step:main_alg3} Set $L := A$.
    \item\label{step:main_alg4} Repeat
    \begin{enumerate}[label = \textbf{(\alph*)}]
        \item\label{step:adding_new} for every $M$ in $J_1, \ldots, J_N$ and row $r$ of $L$, if $r M$ does not belong to the row space of $L$, append $rM$ to $L$.
        \item if nothing has been appended in the previous step, exit the repeat loop and go to~\ref{step:return}.
    \end{enumerate}
    \item\label{step:return} Return $L$.
\end{enumerate}
\end{algorithm}

Let $J(\mathbf{x})$ be the Jacobian matrix of $\mathbf{f}$. From~\citep{LiRabitz}, $L$ gives a lumping of~\eqref{eq:main} if and only if
the row space of $L J(\mathbf{x})$ 
is contained in the row space of $L$ \emph{for all} $\mathbf{x}$.
The universal quantifier in this characterization can be handled in different ways (e.g., see~\citep[\S 3]{LiRabitz} and~\citep[pages~722-723]{Brochot2005}).
We  eliminate it as follows.
Since the entries of $J(\mathbf{x})$ are polynomials in $\mathbf{x}$, we can write $J(\mathbf{x})$ as
\begin{equation}\label{eq:Ji}
  J(\mathbf{x}) = \sum\limits_{i = 1}^N J_i m_i,
\end{equation}
where $m_1, \ldots, m_N$ are distinct monomials in $\mathbf{x}$ and $J_1, \ldots, J_N$ are matrices over $\mathbb{R}$.
Then, the fact that the row space of $L J(\mathbf{x})$ is contained in the row space of $L$ for every $\mathbf{x}$ is equivalent to the containment of  the row space of $L J_i$  
in the row space of $L$ for every $i = 1, \ldots, N$ (proved in the supplementary material, see Lemma~I.1; the equivalence does not hold for the method from~\cite[\S 3]{LiRabitz}, see Remark~I.1).
This leads to the following algorithm: we start with matrix $A$ and add products of its rows with the matrices $J_1, \ldots, J_N$ as long as the dimension of the row space grows.
This is detailed in Algorithm~\ref{alg:simplified}.

\begin{example}
We  illustrate Algorithm~\ref{alg:simplified} by applying it to the system in Eq.~\eqref{eq:ex_running} by choosing $A = (1, 0, 0)$ (thus corresponding to recovering $x_1$ in the reduced system).
The Jacobian matrix of $\mathbf{f}(\mathbf{x})$ in Eq.~\eqref{eq:ex_running} is
    \[
    J(\mathbf{x}) = 
    \begin{pmatrix}
      0 & 2x_2 + 4x_3 & 4x_2 + 8 x_3\\
      -2 & 0 & 4\\
      1 & 1 & 0
    \end{pmatrix},
    \]
which can be decomposed as $J_1m_1 + J_2m_2 + J_3m_3$, where $m_1 = 1$, $m_2 = x_2$, $m_3 = x_3$, and
    \begin{equation*}
      J_1 = \begin{pmatrix}
        0 & 0 & 0\\
        -2 & 0 & 4 \\
        1 & 1 & 0
      \end{pmatrix},\quad
      J_2 = \begin{pmatrix}
        0 & 2 & 4 \\
        0 & 0 & 0 \\
        0 & 0 & 0
      \end{pmatrix},\quad
      J_3 = \begin{pmatrix}
        0 & 4 & 8 \\
        0 & 0 & 0\\
        0 & 0 & 0
      \end{pmatrix}.
  \end{equation*}
  Starting with $L = (1, 0, 0)$, for $r = (1, 0, 0)$ we compute the
  products:
  \[
  rJ_1 = (0, 0, 0), \quad rJ_2 = (0, 2, 4), \quad
  rJ_3 = (0, 4, 8).
  \]
  Since $rJ_1$ belongs to the row space of $L$ while $rJ_2$ does not,
  we set
  \[
  L = \begin{pmatrix}
  1 & 0 & 0\\
  0 & 2 & 4
  \end{pmatrix}.
  \]
  The third vector, $rJ_3$, is proportional to the second row of the new $L$, so we skip it.
  Since a new row, $(0, 2, 4)$, has been added in part~\ref{step:adding_new} of~\ref{step:main_alg4}, we do not exit the loop.
  Setting now $r = (0, 2, 4)$, we get
  \[
    rJ_1 = (0, 4, 8), \quad rJ_2 = (0, 0, 0), \quad rJ_3 = (0, 0, 0).
  \]
  Since all these vectors belong to the row space of $L$, the iteration terminates and the as-computed $L$ gives the lumping of the smallest dimension from which we can recover the original quantities specified through $A$. 
  This $L$ is not equal to the one in~\eqref{eq:lumping_matrix} but is equivalent to it in the above sense.
\end{example}

%%%%%%%%%%%%%%%%%%%%%%%%%%%%%%%%%%%%%%%%%%%%%%%%%%%%%%%%%%%

\section{Implementation}
\begin{algorithm}
\caption{{Finding the smallest invariant subspace\\
(to be used instead of~\ref{step:main_alg4} of Algorithm~\ref{alg:simplified})}}\label{alg:inv_subspace}
\begin{description}[labelsep=2pt]
\item[Input] an $s \times n$ matrix $A$ over field $\mathbb{K}$ and a list $M_1, \ldots, M_\ell$ of $n \times n$ matrices over $\mathbb{K}$;
\item[Output] an $r \times n$ matrix $L$ over $\mathbb{K}$ such that
\begin{itemize}[labelsep=2pt, leftmargin=2mm]
    \item the row span of $A$ is contained in the row span of $L$.
    \item for every $1 \leqslant i \leqslant \ell$, the row span of span of $LM_i$ is contained in the row span of $L$;
    \item $r$ is the smallest possible.
\end{itemize}
\vspace{-0.07in}
\end{description}

\begin{enumerate}[label = \textbf{(Step~\arabic*)}, leftmargin=*, align=left, labelsep=2pt]
    \item Let $L$ be the reduced row echelon form of $A$.
    \item Set $P$ be the set of indices of the pivot columns of $L$.
    \item\label{step:while} While $P \neq \varnothing$ do
    \begin{enumerate}[leftmargin=0mm, label=(\alph*)]
        \item For every $j \in P$ and every $1 \leqslant i \leqslant \ell$
        \begin{enumerate}[leftmargin=4mm]
            \item Let $r$ be the row in $L$ with the index of the pivot being $j$.
            \item Reduce $r M_i$ with respect to $L$. If the result is not zero, append it as a new row to $L$.
            \item Reduce other rows with respect the new one in order to bring $L$ into the reduced row echelon form.
        \end{enumerate}
        \item Let $\widetilde{P}$ be the set of indices of the pivot columns of $L$.
        \item Set $P := \widetilde{P} \setminus P$.
    \end{enumerate}
    \item Return $L$.
\end{enumerate}
\end{algorithm}

The CLUE algorithm was implemented in Python using the SymPy library~\citep{SymPy}. The source code and all  examples from Section~\ref{sec:examples} are available at~\href{https://github.com/pogudingleb/CLUE}{https://github.com/pogudingleb/CLUE}.
{Our implementation accepts models typed manually or provided in the input format of the tool ERODE~\citep{tacas17} (which has an importer from SBML, see~\citet{SBML} for details).}

For our implementation, we keep the general framework of Algorithm~\ref{alg:simplified} but replace~\ref{step:main_alg4}, the most time-consuming part,
with a more efficient algorithm.
\ref{step:main_alg4} of Algorithm~\ref{alg:simplified} solves the following problem: given a set of $n$-dimensional vectors and a set of $n \times n$ matrices, find a basis of the smallest vector space that is invariant under the matrices and that contains the vectors.

We present and implement two algorithms, Algorithm~\ref{alg:inv_subspace} and~\ref{alg:modular} 
for replacing \ref{step:main_alg4} of Algorithm~\ref{alg:simplified} {(we apply them with $A = L$, $\ell = N$, and $M_i = J_i$ for every $i$).
The former is typically faster if the intermediate results of the computation are not too large, while the performance of the latter is less sensitive to the intermediate expression swell, and it can tackle models that are out of reach for the former (for  more details, see Section V in the Supplementary Materials).
Algorithm~\ref{alg:modular} requires the system to have rational coefficients, but all of the systems we  considered are of this type. 
Therefore, our implementation runs Algorithm~\ref{alg:inv_subspace} first and, if the size of intermediate expressions exceeds a threshold (10000 digits), stops it and runs Algorithm~\ref{alg:modular}.
We now describe the key ideas behind these algorithms.}

\textbf{Algorithm~\ref{alg:inv_subspace}} is a result of applying the following observations to~\ref{step:main_alg4} of Algorithm~\ref{alg:simplified}:
\begin{itemize}[labelsep=2pt,itemsep=0.01cm]
    \item If \emph{we maintain $L$ in the reduced row echelon form}, we can test whether a vector $rM$ belongs to the row space of $L$  in $\mathrm{O}(n^2)$ instead of computing rank,
    e.g., in $\mathrm{O}(n^3)$ using Gaussian elimination or in $\mathrm{O}(n^{2.373})$ using more advanced algorithms~\cite[Section~16.5]{ac}.
    \item \emph{We do not need to consider all the products} of the from $rM$ but only the ones corresponding to the pivots of $L$ added at the previous iteration of the loop. The largest number of products considered is reduced from about $n^2N/2$ 
    to at most $nN$; for justification and details, see the proof of Proposition~I.1 in the Supplementary Materials.
\end{itemize}
In  CLUE, we also take advantage of the fact that the input matrices are frequently
very sparse {(see Remark III.1 in the Supplementary Materials)}. 

\begin{algorithm}
\caption{Finding the smallest invariant subspace (modular)\\
(to be used instead of~\ref{step:main_alg4} of Algorithm~\ref{alg:simplified})}\label{alg:modular}
\begin{description}[labelsep=2pt]
\item[Input]  
$s \times n$ matrix $A$ 
and a list $M_1, \ldots, M_\ell$ of $n \times n$ matrices over $\mathbb{Q}$;
\item[Output] an $r \times n$ matrix $L$ over $\mathbb{Q}$ such that:
\begin{itemize}[labelsep=2pt, leftmargin=2mm]
    \item the row span of $A$ is contained in the row span of $L$.
    \item for every $1 \leqslant i \leqslant \ell$, the row span of
    $LM_i$ is contained in the row span of $L$;
    \item $r$ is the smallest possible.
\end{itemize}
\end{description}
\begin{enumerate}[label = \textbf{(Step~\arabic*)}, leftmargin=*, align=left, labelsep=2pt]
    \item\label{step:repeat_loop} Repeat the following
    \begin{enumerate}[leftmargin=0mm, label=(\alph*)]
        \item Pick a prime number $p$ that does not divide any of the denominators in $A, M_1, \ldots, M_\ell$ and has not been chosen before.
        \item 
        Compute the  reductions $\widetilde{A}, \widetilde{M}_1, \ldots, \widetilde{M}_\ell$ modulo $p$.
        \item Run Algorithm~\ref{alg:inv_subspace} on $\widetilde{A}, \widetilde{M}_1, \ldots, \widetilde{M}_\ell$ as matrices over $\mathbb{F}_p$ and denote the result by $\widetilde{L}$.
        \item\label{step:repeat} Apply the rational reconstruction algorithm~(\citep[\S~5.10]{MCA}, \citep{RatRecon}) to construct a matrix $L$ over $\mathbb{Q}$ such that the reduction of $L$ mod  
        $p$ 
        equals
        $\widetilde{L}$.
        \item Check whether the row span of $L$ contains the row span of $L$ and is invariant under $M_1, \ldots, M_\ell$.
        If yes, exit the loop.
    \end{enumerate}
    \item Return the matrix $L$ from
    step~\ref{step:repeat} of the last iteration of the loop.
\end{enumerate}
\end{algorithm}

{As  mentioned, Algorithm~\ref{alg:inv_subspace} may not finish in reasonable time if the intermediate results of computation (exact rational numbers) grow too large (for examples, see Table~2 in the Supplementary Materials).}
We overcome this difficulty in \textbf{Algorithm~\ref{alg:modular}} 
as follows.
We run  Algorithm~\ref{alg:inv_subspace} modulo a prime number.
This may be much faster as it replaces computations with
big integers using long arithmetic by computation with residues that typically fit into 64 bits.
Then we reconstruct a possible output over rational numbers using rational reconstruction 
(see~\cite[\S~5.10]{MCA}) and verify its correctness.
Since the integers in the output (unlike the ones  in the computation) are typically small, the residue modulo prime usually contains already enough information for correct reconstruction.
This verification stage is fast because it deals with the final output that is typically sparse and involves only small integers.
We proceed with the a larger prime until the output is correct.
In all practical examples we considered, one prime  was
enough (we start with $2^{31} - 1$).

The correctness of Algorithms~\ref{alg:inv_subspace} and~\ref{alg:modular} is proved in 
Propositions~I.1 and~I.3 in 
Supplementary Materials.
{We perform a complexity analysis of Algorithms~\ref{alg:simplified} and~\ref{alg:inv_subspace} in Section~III of Supplementary Materials.}

%%%%%%%%%%%%%%%%%%%%%%%%%%%%%%%%%%%%%%%%%%%%%%%%%%%%%%%%

\section{Examples}\label{sec:examples}

In this section, we show the applicability of
CLUE to the reduction of biological models through a number of case studies published in the literature, including some taken from the BioModels repository~\citep{BioModels2010}. We additionally compare CLUE
against the forward equivalence from~\citep{pnas17}. Forward equivalence identifies a partition of an ODE system with polynomial derivatives which induces a lumping where each macro-variable is equal to the sum of variables in each partition block. Using established terminology for lumping methods~\citep{Wei:1969aa,okino1998,Snowden:2017aa}, forward equivalence can be understood as a form of \emph{proper lumping} because each original variable contributes exactly to one macro-variable of the reduced system. By contrast, in general, \cl\ 
yields an \emph{improper} lumping matrix because the linear combination can be arbitrary. Similarly to \cl, 
for forward equivalence there exists the notion of coarsest partition. This is the partition with the smallest number of blocks, thus leading to a reduced ODE system of the smallest dimension. In addition, forward equivalence can be computed with respect to constraints, which are encoded as an initial partition of variables. The algorithm for computing the coarsest partition iteratively splits each block of the initial partition until the criteria for forward equivalence are satisfied. For the comparison, we used ERODE~\citep{tacas17}, which implements the reduction algorithm for forward equivalence. To the best of our knowledge, ERODE is the only publicly available software tool that supports exact lumping for polynomial differential equations. For each case study, both CLUE and forward equivalence were initialized so as to preserve the same observables in the reduced models.

For this study, we computed reductions that were independent from the specific choice of the kinetic parameters used in the models. This was done as follows. Given the original model in the form $\dot{\mathbf{x}} = \mathbf{f}(\mathbf{x}, \mathbf{k})$, where $\mathbf{k}$ is the vector of mass-action kinetic parameters, we considered an extended ODE system with the additional set of equations $\dot{\mathbf{k}} = 0$.  This ensures that the reduction is independent from the initial conditions of the extended variables, hence of the choice of the original parameters.

\begin{table}
\caption{Results for the case studies in Section~\ref{sec:examples} comparing 
CLUE
with forward equivalence (FE) obtained using ERODE for the reduction of the state variables (Vars) and the kinetic parameters (Params). 
}\label{tab:results}
\begin{center}
\begin{tabular}{l|rrr|rrr}
\toprule
& \multicolumn{3}{c|}{\emph{Vars}} 
& \multicolumn{3}{c}{\emph{Params}}\\
\cmidrule(l){2-4} \cmidrule(l){5-7} 
\multicolumn{1}{c|}{\emph{Model}} & \multicolumn{1}{c}{\emph{Orig.}} &
\multicolumn{1}{c}{\emph{~FE~}} & \multicolumn{1}{c|}{\emph{CLUE}} & 
\multicolumn{1}{c}{\emph{Orig.}} &  \multicolumn{1}{c}{\emph{~FE~}} &
\multicolumn{1}{c}{\emph{CLUE}}\\
\midrule
\citet{li2006stochastic} & $21$ & $19$ & $15$ & $25$ & $22$ & $22$ \\
\midrule
\citet{sneddon2011efficient} & & & &  &  & \\
\multicolumn{1}{c|}{$m = 2$}  & $18$ & $12$ & $6$ & $6$ & $6$ & $6$\\ 
\multicolumn{1}{c|}{$m = 3$}  & $66$ & $22$ & $6$ & $6$ & $6$ & $6$\\ 
\multicolumn{1}{c|}{$m = 4$}  & $258$ & $37$ & $6$ & $6$ & $6$ & $6$\\ 
\multicolumn{1}{c|}{$m = 5$}  & $1026$ & $58$ & $6$ & $6$ & $6$ & $6$\\ 
\multicolumn{1}{c|}{$m = 6$}  & $4098$ & $86$ & $6$ & $6$ & $6$ & $6$\\ 
\multicolumn{1}{c|}{$m = 7$}  & $16386$ & $122$ & $6$ & $6$ & $6$ & $6$ \\ 
\multicolumn{1}{c|}{{$m = 8$}}  & {$65538$} & {$167$} & {$6$} & {$6$} & {$6$} & {$6$} \\
\midrule
\citet{Faeder01042003} & {$354$} & {$105$} & {$75$} & {$22$} & {$22$} & {$9$} \\
\midrule
\citet{borisov2008domain} & $213$ & $66$ & $4$ & $14$ & $10$ & $2$ \\ 
\bottomrule
\end{tabular}
\end{center}
\end{table}

\begin{table}
\caption{Runtimes (in seconds) for the numerical solutions of the original model (third column) and the reduced models by forward equivalence (FE) and CLUE using initial conditions as in the original publications and time horizon in column ($H$); columns \emph{Red.} indicate the runtimes for the respective reduction algorithms. Entries \emph{t/o} indicates timeout after 6 hours. }\label{tab:runtimes}
\begin{center}
\begin{tabular}{lrr|rr|rr}
\toprule
& & & \multicolumn{2}{c|}{\emph{FE}} 
& \multicolumn{2}{c}{\emph{CLUE}}\\
\cmidrule(l){4-5} \cmidrule(l){6-7} 
\multicolumn{1}{c}{\emph{Model}} & \multicolumn{1}{c}{$H$} & \multicolumn{1}{c|}{\emph{Sim.}} & \multicolumn{1}{c}{\emph{Red.}} &
\multicolumn{1}{c|}{\emph{Sim.}} &  \multicolumn{1}{c}{\emph{Red.}} &
\multicolumn{1}{c}{\emph{Sim.}}\\
\midrule
\citet{li2006stochastic} & 100 & 0.004 & 0.001 & 0.003 & 0.023 & 0.003 \\
\midrule
\citet{sneddon2011efficient} & 3 & & & & \\
\multicolumn{1}{c}{$m = 2$} & & 0.006 & 0.001 & 0.004 & 0.012  & 0.003 \\
\multicolumn{1}{c}{$m = 3$} & & 0.021 & 0.002 & 0.007 & 0.048 & 0.003 \\
\multicolumn{1}{c}{$m = 4$} & & 0.534 & 0.008 & 0.010 & 0.245 & 0.003 \\
\multicolumn{1}{c}{$m = 5$} & & 20.9 & 0.035 & 0.032 & 1.115 & 0.003 \\
\multicolumn{1}{c}{$m = 6$} & & 1125.0 & 0.185 & 0.299 & 5.238 & 0.005 \\
\multicolumn{1}{c}{$m = 7$} & & \multicolumn{1}{c|}{t/o} & 1.103 & 1.266 & 23.460 & 0.005 \\
\multicolumn{1}{c}{$m = 8$} & & \multicolumn{1}{c|}{t/o} & 3.552 & 6.128 & 97.509 & 0.005 \\
\midrule
\citet{Faeder01042003} & 100 & 0.268 & 0.110 & 0.035 & 0.863 & 0.023 \\
\midrule
\citet{borisov2008domain} & 50 & 0.065 & 0.006 & 0.010 & 0.263  & $<$0.001 \\ 
\bottomrule
\end{tabular}
\end{center}
\end{table}
Tables~\ref{tab:results} and~\ref{tab:runtimes} summarize the numerical results for the models that are analyzed in more detail in the next subsections. For each model, Table~\ref{tab:results} reports the reductions in both the state variables and the number of kinetic parameters.
{Overall, CLUE provides better reductions than ERODE because the partition found by forward equivalence is not guaranteed to be the minimal constrained linear lumping in general. Table~\ref{tab:runtimes} compares the runtimes of the reduction algorithms and of the numerical ODE solutions of the reduced models, computed on commodity hardware. CLUE requires more computation time.
It must be noted that a precise comparison of the reduction runtimes is difficult because the forward equivalence and CLUE are implemented in different programming languages (i.e. Java and Python, respectively); instead, to allow for a fair comparison between the runtimes of the numerical ODE solutions, the reduced models were converted and solved in Matlab (using the stiff ODE solver \texttt{ode15s}).} 

{ The results indicate that the availability of a reduction becomes more relevant as the size of the original model grows. In particular, it makes feasible the analysis of models that would otherwise take considerable time (i.e., more than six hours in the cases $m > 7$ in the model by~\cite{sneddon2011efficient}. 
Nevertheless, we remark that the reduction can be still useful if the runtimes of the reduction algorithms are of the same order as the runtimes of the numerical ODE solutions, especially in cases in
which the reduced model must be subjected to an analysis involving multiple simulations, e.g. to study sensitivity to parameter changes or to initial conditions. }

\subsection{Modular decomposition of signaling pathways}\label{sec:mod_dec}

\begin{figure}
\centering
\includegraphics[width=.9\linewidth]{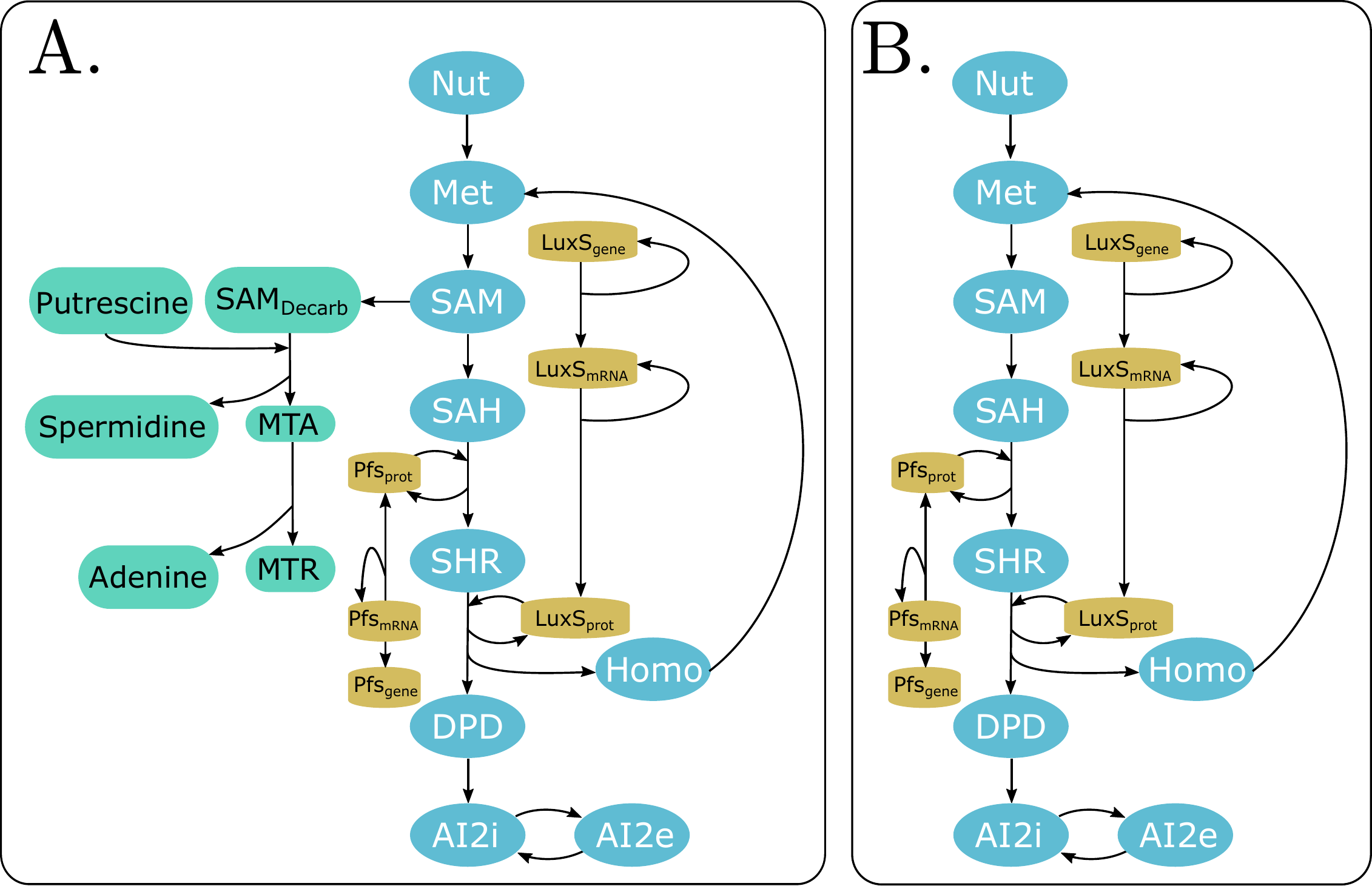}
\caption{(A) Model for AI-2 synthesis adapted from~\citep{li2006stochastic}. (B) Reduced network obtained with CLUE.}\label{fig:QS}
\end{figure}
We first illustrate how CLUE can decompose models of signaling pathways if only certain observables of interest are chosen.  Figure~\ref{fig:QS}(A) depicts a quorum sensing network for \ce{AI 2} biosynthesis and uptake pathways in \emph{E. coli}~\citep{li2006stochastic}; the biochemical model is available in the BioModels repository as \biomodel{MODEL8262229752}. The substrate \ce{Methionine} (\ce{Met}) transforms into S-adenosylmethionine (\ce{SAM}). The blue  branch of the pathway is involved in the production of \ce{AI 2}. The green branch depicts decarboxylation of \ce{SAM}, which ultimately produces \ce{MTR} and \ce{Adenine}. The dynamics of both branches are mediated by \ce{Pfs}. 

The original model has 21 variables,  one for each biochemical species depicted in the pathway. By fixing the output signal \ce{AI 2} as the variable to be preserved,  
CLUE
reduces the system  to 15 variables. An inspection of the reduced model reveals that CLUE
removes the biochemical species depicted as green boxes in Fig.~\ref{fig:QS}(A), while the remaining variables are not aggregated further.  Overall, this leads to a reduced model that can be interpreted as the network in Fig.~\ref{fig:QS}(B). The reduction can be explained by the fact that none of the eliminated variables contribute to the dynamics of the chosen observables. This is because the interactions between the green pathway and the blue one occur only through  \ce{Pfs}, which acts a catalyst in all the reactions in which it is involved. 

The largest reduction by forward equivalence 
that preserves AI 2 has 19 variables. It
only aggregates Adenine, MTR, and Spermidine in the same block, while keeping all the other variables separated. This reduction is, however, a trivial one because these are end products of the pathways that do not interact with any other  species. Mathematically, this results in the differential equation of an end-product variable not featuring the variable itself in the right-hand side. As a consequence, end-product variables can always be rewritten in terms of a lumped variable that represents their sum. A similar pattern of modular decomposition, on a pathway model of cartilage breakdown from~\citep{proctor2014computer}, is discussed in Supplementary Materials. 

%%%%%
\subsection{Multisite protein phosphorylation}\label{CS:e3}
Here we study a basic mechanism of protein phosphorylation, a fundamental process in eukaryotic cells~\citep{Gunawardena11102005}, to show how CLUE
can help cope with the combinatorial growth of mechanistic models for proteins with multiple sites~\citep{FEBS:FEBS7027}.  We consider a model phosphorylation/dephosphorylation of a substrate with $m$ independent and identical binding sites, taken from~\citep{sneddon2011efficient}. Each site can be in four different states: phosphorylated and unbound, unphosphorylated and unbound, phosphorylated and bound to a phosphatase, unphosphorylated and bound to a kinase. Thus, the model is described by $4^m+2$ variables to track all possible protein configurations, in addition to the concentrations of the free kinase and phosphatase.

\begin{figure}[t]
\centering
\includegraphics[width=.9\linewidth]{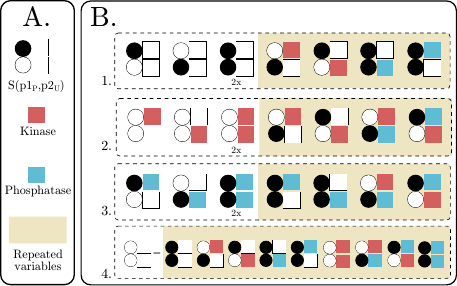}
\caption{Application of CLUE to 
a protein phosphorylation model with $m=2$ identical and independent binding sites. A) Model components: empty/full circles denote whether the site is unphosphorylated/phosphorylated; binding of a kinase or phosphatates is denoted by the color of the square. B) Graphical representation of the four macro-variables obtained by CLUE; the yellow background groups variables that appear in more than one macro-variable; we write `2x' under variables that are counted twice.}\label{fig:e2}
\end{figure}

 For $m=2$ independent sites, CLUE
reduces the model from 18 to 6 variables if
observing the free kinase (or the free phosphatase). In the reduced model, 2
macro-variables represent the free kinase and phosphatase, respectively. The other
macro-variables are linear combinations of the protein configuration (Fig.~\ref{fig:e2}). An inspection of the aggregation shows that 3
of these macro-variables represent the total concentration of a specific binding-site configuration: 
free and phosphorylated (Fig.~\ref{fig:e2}-B1); free and bound to a kinase (Fig.~\ref{fig:e2}-B2); phosphorylated and bound to a phosphatase (Fig.~\ref{fig:e2}-B3). Thus,
if the two binding sites have the same configuration, the corresponding variable is counted twice.
Also, the aggregation results in an improper lumping
as there are variables that contribute to more than one macro-variable. The last macro-variable  escaped physical intelligibility  as
it represents the difference between the free substrate with unphosphorylated sites and protein configurations that appear in the aforementioned lumps. 
Interestingly, running CLUE for models with more binding sites until $m={8}$ always returned a 6-dimensional reduced model that obey the patterns similar to the one from Fig.~\ref{fig:e2}.  
Instead, forward equivalence detects the assumption of the binding sites being identical~\citep[Table S1]{pnas17}. 
Each macro-variable represents complexes equal up to  
permutation of the identities of the binding sites, leading to a
polynomial growth in 
$m$ of the number of variables in the reduced model.

 {A different type of aggregation in multisite phosphorylation models is discussed in~\citep{pnas17} regarding the rule-based model of early events of the signaling pathway of the high-affinity receptor for IgE (Fc$\varepsilon$RI) in~\citep{Faeder01042003}. In this model, the bivalent IgE ligand aggregates Fc$\varepsilon$RI through phosphorylation of the tyrosine residues on its $\beta$ and $\gamma$ subunits by the Lyn kinase, which in turn recruits the tyrosine kinase Syk by phosphorylation of two distinct units. The original model consists of 354 variables, each representing a distinct biochemical complex. In~\citep{pnas17}, the maximal forward equivalence is reported to yield a reduced model with 105 variables in which all complexes that have the same configuration except the phosphorylation state of the Syk units are lumped into the same equivalence class. This model can be further aggregated by CLUE. Observing the concentration of the ligand-receptor complex when both $\beta$ and $\gamma$ binding sites are phosphorylated gives a model with 84 macrovariables. This reduction is coarser than the largest forward equivalence because it contains macrovariables that represent linear combinations with negative coefficients or with positive coefficients different than one. Because of this, the reduction cannot be found using specific methods for rule-based systems~\citep{Borisov2005951,16430778,Feret21042009}. An inspection of the lumping matrix reveals macrovariables that may carry a physical interpretation. Specifically, it is possible to identify three classes of aggregation. Each class consists of macrovariables that are sums of three original variables, each representing a biomolecular complex consisting of both receptors bound to the bivalent ligand. Across original variables within a macrovariable, one receptor is found in each of the following forms: bound to Syk, phosphorylated at the $\gamma$ subunit, or unphosphorylated at the $\gamma$ subunit. The three classes of aggregation are characterized by the state of the other receptor, which is the same within each macrovariable. In particular, there are 7 macrovariables in which the other receptor has a phosphorylated and unbound $\gamma$ site; 3 macrovariables in which the $\gamma$ site is phosphorylated and bound to Syk; and 6 macrovariables in which the $\gamma$ site is unbound and unphosphorylated.}

\begin{figure}[t]
\centering
\includegraphics[width=.9\linewidth]{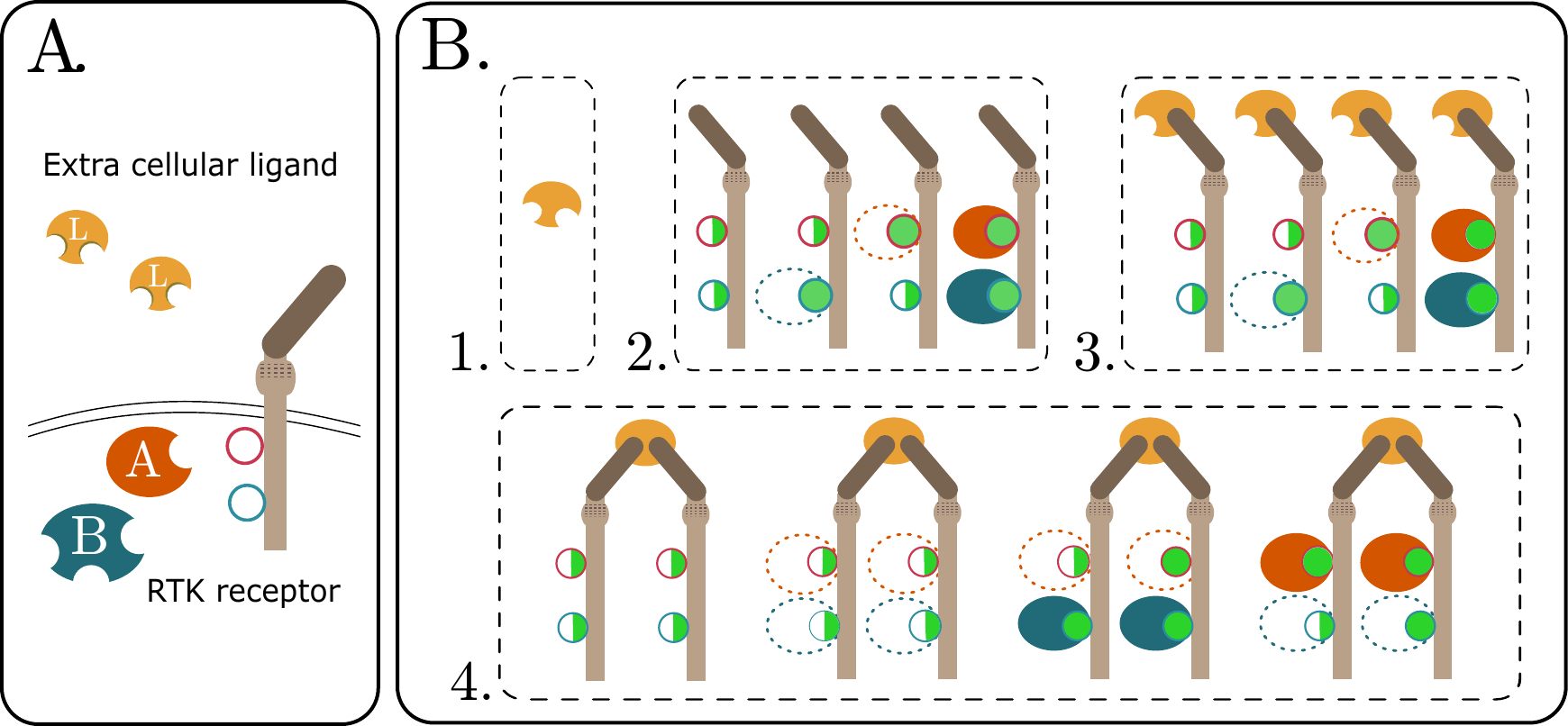}
\caption{(A) Components in a model for RTK signaling 
adapted from \citep{borisov2008domain}: a bivalent Ligand (L), 2 adapter proteins A and B and a receptor with 3 binding sites: ligand binding site in the extracellular region (brown); protein binding sites in the intracellular region (red/blue circles). (B) macro-variables obtained in the reduced model.}\label{fig:m7}
\end{figure}

\subsection{Aggregation for ordered phosphorylation mechanisms}\label{sec:rtk} 

We now consider an example of ordered phosphorylation, taken from~\citep{borisov2008domain}, in a receptor tyrosine kinase (RTK) signaling pathway where receptor autophosphorylation via dimerization is preceded by ligand binding. Figure~\ref{fig:m7}(A) shows the molecular complexes involved in the pathway.  The receptor interacts with a bivalent ligand and two adapter proteins, A and B. Protein \ce{A} has a single site that binds to the receptor. Protein \ce{B} is a scaffold protein with three binding sites: one extracellular site dedicated to receptor-binding and two intracellular tyrosine residues. The phosphorylation state of the tyrosine residues in \ce{B} is independent of the state of the receptor-binding site.
Upon phosphorylation of the intracellular sites, the receptor can bind the adapter proteins \ce{A} and \ce{B}. 

The model, originally expressed in the rule-based language BioNetGen~\citep{Blinov22112004}, has 213 variables. 
Applying CLUE
to preserve the concentration of free ligand yields a reduced model where 150 variables are removed and the remaining 63 are lumped into four macro-variables, depicted in Fig.~\ref{fig:m7}-B. These represent: the free ligand (Fig.~\ref{fig:m7}-B1); all configurations of the free receptor regardless of the phosphorylation state of the intracellular terminals or of protein binding (Fig.~\ref{fig:m7}-B2); all variables that represent the bound ligand (Fig.~\ref{fig:m7}-B3) and the dimerized form (Fig.~\ref{fig:m7}-B4) regardless of the intracellular states.  Instead,  forward equivalence gives 66 variables, aggregating B-bound receptor units regardless of the phosphorylation state of \ce{B}. 

%%%%%%%%%%%%%%%%%%%%%%%%%%%%%%%%%%%%%%%%%%%%%%%%%%%%%%%%%

\section{Conclusion}
We presented CLUE, an algorithm for 
the reduction of polynomial 
ODEs by exact lumping, with the possibility to fix original variables (or their linear combinations)
to be recovered in the reduced system. From a practical viewpoint, the specification of such constraints allows the preservation of the dynamics of key biochemical species of interest to the modeler. Importantly, although it is acknowledged that linear lumping may lead to loss of structure in the reduced model, e.g.,~\citep{Snowden:2017aa}, the reductions presented here admitted a biochemical interpretation in most cases. From a computational viewpoint, 
CLUE
casts the analysis of polynomial equations into a linear-algebra framework, allowing reductions for models of dimension over than $15{,}000$ variables using a prototype implementation. This makes CLUE a general-purpose tool that adds to the wide range of existing methods.  
In particular, since it
reduces exactly, it can be used as a pre-processing for techniques that seek more aggressive reductions using approximate methods, or as a complementary method to those that use orthogonal model properties, e.g. time-scale separation.

\section*{Funding}

This work has been supported by the National Science Foundation [CCF-1563942 and 1564132, DMS-1760448, 1853650, and 1853482], by the MIUR PRIN project ``SEDUCE'', no. 2017TWRCNB, and by the Paris Ile-de-France Region.

\bibliographystyle{abbrvnat}
\bibliography{references}

\end{document}

\maketitle

\setcounter{lemma}{0}
\renewcommand{\thelemma}{I.\arabic{lemma}}
\setcounter{proposition}{0}
\renewcommand{\theproposition}{I.\arabic{proposition}}
\setcounter{notation}{0}
\renewcommand{\thenotation}{I.\arabic{notation}}
\setcounter{remark}{0}
\renewcommand{\theremark}{I.\arabic{remark}}
\vspace{-0.3in}
This document is structured as follows:
\begin{itemize}
    \item In Section~\ref{sec:Alg3}, we will prove the correctness and termination the algorithms presented in the paper.
    \item In Section~\ref{sec:LiRabitzproof}, we reprove the criterion for lumping in terms of the Jacobian of the system~\cite[Section~2]{LI199195} for the sake of completeness.
    \item In Section~\ref{sec:complexityanalysis}, we present a complexity analysis of our algorithms and compare it with the complexity of ERODE.
    \item In Section~\ref{sec:cartilage}, we discuss the application of CLUE to a cartilage breakdown model in~\citep{proctor2014computer}.
    \item In Section~\ref{sec:runtimes}, we {compare the performance of our implementations Algorithms~\ref{alg:inv_subspace} and Algorithm~\ref{alg:modular}.}
 \end{itemize}

\begin{remark*}
{
In the present paper, we have focused on exact maximal reduction for ODEs with a polynomial right-hand side because our examples are in this class. However, all of our algorithms can be immediately applied to other kinds of systems, such as discrete-time polynomial systems (e.g. $\mathbf{x}_{n+1} = \mathbf{f}(\mathbf{x}_n)$).
}
\end{remark*}

\section{Proofs of correctness and termination of algorithms}\label{sec:Alg3}
For the convenience of the reader while navigating between the main paper and the Supplementary materials, we recall:

\begin{algorithm}[H]
\caption{Simplified algorithm for finding a constrained lumping of the smallest possible dimension}\label{alg:simplified}
\begin{description}[itemsep=0pt]
\item[Input ] a system $\dot{\mathbf{x}} = \mathbf{f}(\mathbf{x})$ of $n$ ODEs with a polynomial right-hand side and an $s \times n$ matrix $A$ over field $\mathbb{K}$ of rank $s > 0$;
\item[Output ] a matrix $L$ such that $\mathbf{y} := L\mathbf{x}$ is a constrained lumping with observables $A\mathbf{x}$ of smallest possible dimension.
\end{description}

\begin{enumerate}[label = \textbf{(Step~\arabic*)}, leftmargin=*, align=left, labelsep=2pt, itemsep=0pt]
    \item\label{step:main_alg1} Compute $J(\mathbf{x})$, the Jacobian matrix of $\mathbf{f}(\mathbf{x})$.
    \item\label{step:main_alg2} Represent $J(\mathbf{x})$ as $J_1m_1 + \ldots + J_Nm_N$, where $m_1, \ldots, m_N$ are distinct monomials in $\mathbf{x}$, and $J_1, \ldots, J_N$ are nonzero matrices over $\mathbb{R}$.
    \item\label{step:main_alg3} Set $L := A$.
    \item\label{step:main_alg4} Repeat
    \begin{enumerate}[label = \textbf{(\alph*)}]
        \item\label{step:adding_new} for every $M$ in $J_1, \ldots, J_N$ and row $r$ of $L$, if $r M$ does not belong to the row space of $L$, append $rM$ to $L$.
        \item if nothing has been appended in the previous step, exit the repeat loop and go to~\ref{step:return}.
    \end{enumerate}
    \item\label{step:return} Return $L$.
\end{enumerate}
\end{algorithm}

\begin{algorithm}[H]
\caption{Finding the smallest invariant subspace\\
(to be used instead of~\ref{step:main_alg4} of Algorithm~\ref{alg:simplified} for $A = L$, $\ell = N$, and $M_i = J_i$ for $1 \leqslant i \leqslant \ell$)}\label{alg:inv_subspace}
\begin{description}[itemsep=0pt]
\item[Input] an $s \times n$ matrix $A$ over field $\mathbb{K}$ and a list $M_1, \ldots, M_\ell$ of $n \times n$ matrices over $\mathbb{K}$;
\item[Output] an $r \times n$ matrix $L$ over $\mathbb{K}$ such that
\begin{itemize}[ leftmargin=2mm, itemsep=0pt]
    \item the row span of $A$ is contained in the row span of $L$.
    \item for every $1 \leqslant i \leqslant \ell$, the row span of span of $LM_i$ is contained in the row span of $L$;
    \item $r$ is the smallest possible.
\end{itemize}
\vspace{-0.07in}
\end{description}

\begin{enumerate}[label = \textbf{(Step~\arabic*)}, leftmargin=*, align=left, itemsep=0pt, topsep=0pt]
    \item\label{alg2:reduced_row} Let $L$ be the reduced row echelon form of $A$.
    \item\label{alg2:pivots} Set $P$ be the set of indices of the pivot columns of $L$.
    \item\label{step:while} While $P \neq \varnothing$ do
    \begin{enumerate}[leftmargin=0mm, label=(\alph*), itemsep=0pt, topsep=0pt]
        \item For every $j \in P$ and every $1 \leqslant i \leqslant \ell$
        \begin{enumerate}[leftmargin=4mm]
            \item Let $v$ be the row in $L$ with the index of the pivot being $j$.
            \item Reduce $v M_i$ with respect to $L$. If the result is not zero, append it as a new row to $L$.
            \item Reduce other rows with respect the new one in order to bring $L$ into the reduced row echelon form.
        \end{enumerate}
        \item Let $\widetilde{P}$ be the set of indices of the pivot columns of $L$.
        \item Set $P := \widetilde{P} \setminus P$.
    \end{enumerate}
    \item Return $L$.
\end{enumerate}
\end{algorithm}

%%%%%%%%

\begin{algorithm}[H]
\caption{Finding the smallest invariant subspace (modular)\\
(to be used instead of~\ref{step:main_alg4} of Algorithm~\ref{alg:simplified} for $A = L$, $\ell = N$, and $M_i = J_i$ for $1 \leqslant i \leqslant \ell$)}\label{alg:modular}
\begin{description}[itemsep=0pt]
\item[Input]  
$s \times n$ matrix $A$ 
and a list $M_1, \ldots, M_\ell$ of $n \times n$ matrices over $\mathbb{Q}$;
\item[Output] an $r \times n$ matrix $L$ over $\mathbb{Q}$ such that:
\begin{itemize}[leftmargin=2mm, itemsep=0pt, topsep=0pt]
    \item the row span of $A$ is contained in the row span of $L$.
    \item for every $1 \leqslant i \leqslant \ell$, the row span of
    $LM_i$ is contained in the row span of $L$;
    \item $r$ is the smallest possible.
\end{itemize}
\vspace{-0.1in}
\end{description}
\begin{enumerate}[label = \textbf{(Step~\arabic*)}, leftmargin=*, align=left, itemsep=0pt, topsep=0pt]
    \item\label{step:repeat_loop} Repeat the following
    \begin{enumerate}[leftmargin=0mm, label=(\alph*)]
        \item Pick a prime number $p$ that does not divide any of the denominators in $A, M_1, \ldots, M_\ell$ and has not been chosen before.
        \item 
        Compute the  reductions $\widetilde{A}, \widetilde{M}_1, \ldots, \widetilde{M}_\ell$ modulo $p$.
        \item Run Algorithm~\ref{alg:inv_subspace} on $\widetilde{A}, \widetilde{M}_1, \ldots, \widetilde{M}_\ell$ as matrices over $\mathbb{F}_p$ and denote the result by $\widetilde{L}$.
        \item\label{step:repeat} Apply the rational reconstruction algorithm~(\citep[\S~5.10]{MCA}, \citep{RatRecon}) to construct a matrix $L$ over $\mathbb{Q}$ such that the reduction of $L$ mod  
        $p$ 
        equals
        $\widetilde{L}$.
        \item Check whether the row span of $L$ contains the row span of $L$ and is invariant under $M_1, \ldots, M_\ell$.
        If yes, exit the loop.
    \end{enumerate}
    \item Return the matrix $L$ from
    step~\ref{step:repeat} of the last iteration of the loop.
\end{enumerate}
\end{algorithm}

%%%%%%%%%%%%%%%%%%%%%%%%%%%%%%%%%%%%%%%

\begin{notation}
\begin{itemize}[labelsep=2pt]
    \item[]
    \item $\Mat_{m, n}(\mathbb{K})$  denotes the space of $m \times n$ matrices over a field $\mathbb{K}$.
    \item For  
    $M\in\Mat_{m,n}(\mathbb{K})$, 
     $\rspan_{\mathbb{K}}(M)$  is the row span of $M$ over $\mathbb{K}$.
\end{itemize}
\end{notation}

Lemma~\ref{lem:decomposing_matrix}
is used by Algorithm~\ref{alg:simplified} to pass from the invariance under the Jacobian to the invariance under a finite set of constant matrices.

\begin{lemma}\label{lem:decomposing_matrix}
  Let $M(\mathbf{x}) \in \Mat_{n, n}(\mathbb{K}[\mathbf{x}])$, where $\mathbf{x} = (x_1, \ldots, x_r)$ and $\Char \mathbb{K} = 0$.
  We write $M(\mathbf{x}) = M_1m_1 + \ldots + M_N m_M$ so that $M_1, \ldots, M_N \in \Mat_{n, n}(\mathbb{K})$ and $m_1, \ldots, m_N$ are distinct monomials in $\mathbf{x}$.
  Then, for a vector subspace $V \subset \mathbb{K}^n$, the following are equivalent: 
  \begin{enumerate}[label=(\arabic*),labelsep=2pt]
      \item $V$ is invariant under $M(\mathbf{x}^\ast)$ for every $\mathbf{x}^\ast \in \mathbb{K}^r$;
      \item $V$ is invariant under $M_i$ for every $1 \leqslant i \leqslant N$.
  \end{enumerate}
\end{lemma}

\begin{proof}
  Assume that $V$ is invariant under $M_1, \ldots, M_N$.
  Since, for every $\mathbf{x}^\ast \in \mathbb{K}^r$, $M(\mathbf{x}^\ast)$ is an $\mathbb{K}$-linear combination of $M_1, \ldots, M_N$, $V$ is invariant under $M(\mathbf{x}^\ast)$ as well.
  
  Assume that $V$ is invariant under $M(\mathbf{x}^\ast)$ for every $\mathbf{x}^\ast \in \mathbb{K}^r$.
  Consider $v \in V$.
  Since for all $\mathbf{x}^\ast \in \mathbb{K}^r$, $M(\mathbf{x}^\ast) v \in V$, for every $1 \leqslant i \leqslant r$, $\frac{\partial M}{\partial x_i} (\mathbf{x}^\ast) v\in V$ as well.
  Consider one of $M_1, \ldots, M_N$, say $M_1$.
  Let $m_1 = x_1^{d_1}\ldots x_r^{d_r}$.
  Iterating the argument with derivative, we  
  obtain  
  \[
\forall \mathbf{x}^\ast \in \mathbb{K}^r\ \    \frac{\partial^{d_1 + \ldots + d_r} M}{\partial x_1^{d_1} \ldots \partial x_r^{d_r}} (\mathbf{x}^\ast) v \in V. 
  \]
  Taking $\mathbf{x}^\ast = \mathbf{0}$, we deduce that $M_1v \in V$.
\end{proof}

\begin{remark}\label{rem:comparison_LR}
 A different approach to replacing the Jacobian with a finite set of constant matrices was suggested in~\cite[Sect.~3(A)]{LiRabitz}:
  \begin{enumerate}
      \item Write the Jacobian $J(\mathbf{x}) = \sum a_{ij}(\mathbf{x})E_{ij}$, where $E_{ij}$ is the matrix with one in the $(i, j)$-th cell and zeroes everywhere else;
      \item Combine together summands with proportional $a_{ij}(\mathbf{x})$ obtaining a representation $J(\mathbf{x}) = \sum b_j(\mathbf{x})B_j$ with constant $B_j$;
      \item Return $B_j$'s.
  \end{enumerate}
  Consider the system
  \[
  \begin{cases}
    \dot{x}_1 = (x_2 + x_3)^2 + (x_2 + x_4)^2,\\
    \dot{x}_2 = \dot{x}_3 = \dot{x}_4 = 0
  \end{cases}
  \]
  with the observable $x_1$.
  Then the procedure from~\cite[Section~3(A)]{LiRabitz}  will lead to the following decomposition
  \[
    J(\mathbf{x}) = 2(2x_2 + x_3 + x_4) E_{12} + 2(x_2 + x_3) E_{13} + 2(x_2 + x_4)E_{14}.
  \]
  The smallest subspace containing $(1, 0, 0, 0)$ and right-invariant under $E_{12}, E_{13}, E_{14}$ is the whole space, so this approach will not produce a nontrivial lumping.
  On the other hand, using Lemma~\ref{lem:decomposing_matrix}, we arrive at 
  \[
  J(\mathbf{x}) = 2x_2(2E_{12} + E_{13} + E_{14}) + 2x_3(E_{12} + E_{13}) + 2x_4(E_{12} + E_{14}).
  \]
  The matrices $2E_{12} + E_{13} + E_{14}, E_{12} + E_{13}$, and $E_{12} + E_{14}$ have a common proper invariant subspace containing $(1, 0, 0, 0)$, and this yields a nontrivial lumping:
  \[
  y_1 = x_1,\quad y_2 = x_2 + x_3,\quad y_3 = x_2 + x_4.
  \]
\end{remark}

%%%%%%%%%%%%%%%%%%%%%%%%%%%%%%%%%%%%%%%%%

\begin{proposition}\label{prop:alg_correct}
  Algorithm~\ref{alg:inv_subspace} is correct.
\end{proposition}

\begin{proof}
  Bringing a matrix to the reduced row echelon form does not change the row span, and adding extra rows might only enlarge it, so the row span of the output of Algorithm~\ref{alg:inv_subspace} contains the row span of $A$.
  
  Now we will show that the row span of the output of the algorithm is invariant under $M_1, \ldots, M_N$.
  We denote the values of $L$ and $P$ before the $i$-th iteration of the while loop~\ref{step:while} by $L_i$ and $P_i$, respectively.
  We set $L_0$ and $P$ to be the $0 \times n$ matrix and $\varnothing$, respectively.
  We will show by induction on $k$ that, for every $k \geqslant 0$ and every $1 \leqslant i \leqslant \ell$, we have
  \begin{equation}\label{eq:rspans}
    \rspan_{\mathbb{K}}(L_kM_i) \subset \rspan_{\mathbb{K}}(L_{k + 1}).
  \end{equation}
  The case $k = 0$ is true.
  Assume that the statement is true for all numbers less than some $k > 0$.
  Let $L_{+}$ be the matrix consisting of the rows of $L_{k}$ with the pivot columns in $P_{k}$, and let $L_{-}$ be the matrix consisting of the remaining rows.
  Fix $1 \leqslant i \leqslant \ell$.
  Then $\rspan_{\mathbb{K}}(L_{+}M_i) \subset \rspan_{\mathbb{K}} L_{k + 1}$ because the rows of $L_{+}$ will be processed in the next iteration of the while loop.
 By the construction, $\rspan_{\mathbb{K}} L_{k - 1} \subset \rspan_{\mathbb{K}} L_k$.  
  The rows of $L_{k - 1}$ and $L_{+}$ are linearly independent because they form a (nonreduced) row echelon form after permuting rows and columns.
  Therefore, $\rspan_{\mathbb{K}} L_{k} = \rspan_{\mathbb{K}} L_{+} + \rspan_{\mathbb{K}} L_{k - 1}$. This implies 
  \[
  \rspan_{\mathbb{K}} (L_{-}M_i) \subset \rspan_{\mathbb{K}} (L_{+}M_i) + \rspan_{\mathbb{K}} (L_{k - 1}M_i). 
  \]
  The inductive hypothesis implies that 
  \[\rspan_{\mathbb{K}} (L_{-}M_i) \subset \rspan_{\mathbb{K}} (L_{+}M_i) + \rspan_{\mathbb{K}} L_{k} \subset \rspan_{\mathbb{K}}L_{k + 1}. 
  \]
  Therefore, $\rspan_{\mathbb{K}}(L_kM_i) \subset \rspan_{\mathbb{K}} L_{k + 1}$.
  
  Assume that there were $N$ iterations of the while loop. 
  Then we consider one extra iteration. 
  Since $P = \varnothing$, this iteration will not do anything, so $L_{N + 2} = L_{N + 1}$.
  Therefore, $\rspan_{\mathbb{K}}(L_{N + 1}M_i) \subset \rspan_{\mathbb{K}}(L_{N + 1})$ for every $1 \leqslant i \leqslant \ell$ due to~\eqref{eq:rspans}.
  This implies that $\rspan_{\mathbb{K}}$ of the output of the algorithm is invariant under $M_1, \ldots, M_\ell$.
  
  To prove the minimality of $r$, consider
  $V$, the smallest subspace of $\mathbb{K}^n$ invariant under $M_1, \ldots, M_\ell$ and containing the rows of the input matrix $A$.
  We will show by induction on $i$ that $\rspan_{\mathbb{K}}(L_i) \subset V$.
  Since $\rspan_{\mathbb{K}}(L_1)=\rspan_{\mathbb{K}}A$, 
  $\rspan_{\mathbb{K}}(L_1) \subset V$.
  Assume that the statement is true for some $i \geqslant 1$.
  At the $i$-th iteration of the while loop, we consider vectors of the form $vM_i$, where $v \in \rspan_{\mathbb{K}}(L_i)$.
  Since $v \in V$ and $V$ is $M_i$-invariant, these vectors also belong to $V$. 
  Consequent computation of the row echelon form does not change the row span.
  Hence, the row span of the output is invariant under $M_1, \ldots, M_\ell$ and contained in $V$, so it 
  coincides with $V$.
  This proves the minimality of $r$.
\end{proof}

The following lemma is used in Proposition~\ref{prop:alg_mod_correct} for showing the correctness and termination of Algorithm~\ref{alg:modular}.

\begin{lemma}\label{lem:mod_aux}
  Let $A \in \Mat_{s, n}(\mathbb{Q})$, $M_1, \ldots, M_\ell \in \Mat_{n, n}(\mathbb{Q})$ and $L$  the result of applying Algorithm~\ref{alg:inv_subspace} to these matrices.
  For every prime number $p$ that does not divide the denominators of the entries of $A, M_1, \ldots, M_\ell$, we denote the result of applying Algorithm~\ref{alg:inv_subspace} to the reductions of these matrices modulo~$p$ by $L_p^\ast$.
  Then 
  \begin{enumerate}[label=(\arabic*)]
     \item\label{item:a2:1} for all but finitely many primes, $L^\ast_p$ is equal to  
      $L$ modulo $p$;
     \item\label{item:a2:2} the number of rows in $L_p^\ast$ does not exceed the number of rows in $L$.
  \end{enumerate}
\end{lemma}

\begin{proof}
To show~\ref{item:a2:1}, consider the run of Algorithm~\ref{alg:inv_subspace} on $A, M_1, \ldots, M_\ell$.
  The operations performed with the matrix entries in the algorithm are arithmetic operations and checking for nullity.
  There is a finite list of nonzero rational numbers $q_1, \ldots, q_N$ checked for nullity in the algorithm.
  Consider a prime number $p$ such that the reductions of $q_1, \ldots, q_N$ modulo $p$ are defined and not  zero.
  Since the arithmetic operations commute with reducing modulo $p$ and we have chosen $p$ so that all nullity checks will also commute with reduction modulo $p$, the result of the algorithm modulo $p$, that is $L_p^\ast$, will be equal to the reduction of $L$ modulo $p$.
  
 We now show~\ref{item:a2:1}. The number of rows in $L$ is the dimension of the space generated by the rows of $A$ and their images under all possible products of $M_1, \ldots, M_\ell$.
  Consider the $\infty \times n$ matrix $R$ formed from the  matrices of the form $AX$, where $X$ ranges over all possible products of $M_1, \ldots, M_\ell$, stacked on top of each other.
  Let $R_p$ be the reduction of $R$ modulo $p$.
  For every integer $r$, having rank at most $r$ can be expressed as a system of polynomial conditions in the matrix entries (that is, all $(r + 1) \times (r + 1)$ minors are zero).
  Therefore, $\operatorname{rank} R_p \leqslant \operatorname{rank} R$.
  Since the numbers of rows in $L$ and $L_p^\ast$ are equal to $\operatorname{rank} R$ and $\operatorname{rank} R_p$, respectively, the second part of the lemma is proved.
\end{proof}

%%%%%%%%%%%%%%%

\begin{proposition}\label{prop:alg_mod_correct}
  Algorithm~\ref{alg:modular} is correct and 
  terminates in finite time.
\end{proposition}

\begin{proof}
  First we will show the correctness.
  Consider the output of Algorithm~\ref{alg:modular}, call it $L_0$.
  Since the stopping criterion for the loop in~\ref{step:repeat_loop} is $\rspan_{\mathbb{Q}}(A) \subset \rspan_{\mathbb{Q}}(L_0)$ and the invariance of $\rspan_{\mathbb{Q}}(L_0)$ under $M_1, \ldots, M_\ell$, it remains to prove the minimality of the number of rows in $L_0$.
  Due to Proposition~A.1 from the main paper (correctness of Algorithm~\ref{alg:inv_subspace}), it would be equivalent to show that the number of rows in $L_0$ is equal to the number of rows in the output of Algorithm~\ref{alg:inv_subspace} on $A, M_1, \ldots, M_\ell$, call it $L$.
  The second part of Lemma~\ref{lem:mod_aux} implies that the number of rows of every matrix $\widetilde{L}$ computed in~\ref{step:repeat_loop} does not exceed the number of rows in $L$.
  Then the same is true for $L_0$.
  Since the number of rows in $L$ is the smallest 
  possible, it is the same as the number of rows in $L_0$, so the output of the algorithm will be correct.

  Now we will prove the termination.
  Let $N$ be the maximum of the absolute values of the numerators and denominators of the entries of $L$.
  Consider a prime number $p$ such that $L_p^\ast$ (see Lemma~\ref{lem:mod_aux}) is equal to the reduction of $L$ modulo $p$ and $p > 2N^2$. 
  Then~\citep{RatRecon} and~\cite[Lemma~2]{RatRecon2} imply that the result of rational reconstruction in~\ref{step:repeat} for $\widetilde{L} = L^\ast_p$ will be equal to $L$, so the algorithm will terminate.
  Lemma~\ref{lem:mod_aux}\ref{item:a2:1} implies that all but finitely many primes satisfy the above properties, so the algorithm will 
  reach one of these numbers and terminate.
\end{proof}

%%%%%%%%%%%%%%%%%%%%%%%%%%%%%%%%%%%%%%%%%%%%%%%%%%
\section{Proof for the lumping criterion from~\cite{LiRabitz}}\label{sec:LiRabitzproof}

\setcounter{lemma}{0}
\renewcommand{\thelemma}{II.\arabic{lemma}}
\setcounter{proposition}{0}
\renewcommand{\theproposition}{II.\arabic{proposition}}
\setcounter{notation}{0}
\renewcommand{\thenotation}{II.\arabic{notation}}
\setcounter{remark}{0}
\renewcommand{\theremark}{II.\arabic{remark}}

In Lemma~\ref{lem:poly_secret} and Proposition~\ref{prop:jacobian}, we reprove the criterion for lumping in terms of the Jacobian of the system~\cite[Section~2]{LI199195} for the sake of completeness.

\begin{lemma}\label{lem:poly_secret}
  Let $p(\mathbf{x}) \in \mathbb{R}[\mathbf{x}]$, where $\mathbf{x} = (x_1, \ldots, x_n)$, and $L \in \Mat_{s, n}(\mathbb{R})$.
  Let $V \subset \mathbb{R}^n$ be the orthogonal complement to $\rspan_{\mathbb{R}}(L)$.
  Then $p(\mathbf{x})$ can be written as a polynomial in $L\mathbf{x}$ if and only if $\forall\mathbf{v} \in \mathbb{R}^n$ the operator $D_{\mathbf{v}} := v_1\frac{\partial}{\partial x_1} + \ldots + v_n\frac{\partial}{\partial x_n}$ annihilates $p(\mathbf{x})$. 
\end{lemma}

\begin{proof}
  Denote the rows of $L$ by $\mathbf{r}_1, \ldots, \mathbf{r}_s$.
  Assume that there exists a polynomial $q$ in $y_1, \ldots, y_s$ such that $p(\mathbf{x}) = q(L\mathbf{x})$.
  Then
\[
 \forall \mathbf{v} \in V\quad  D_{\mathbf{v}} p(\mathbf{x}) = D_{\mathbf{v}} q(L\mathbf{x}) = (\mathbf{v}, \mathbf{r}_1) \frac{\partial q}{\partial y_1}(L\mathbf{x}) + \ldots + (\mathbf{v}, \mathbf{r}_s) \frac{\partial q}{\partial y_s}(L\mathbf{x}) = 0.
 \]
  To prove the lemma in the other direction, choose an orthonormal basis $\mathbf{u}_1, \ldots, \mathbf{u}_\ell$ of $V$.
  Since the rows of $L$ and $\mathbf{u}_1, \ldots, \mathbf{u}_\ell$ span the whole space, there exists a polynomial $q$ in $y_1, \ldots, y_{s + \ell}$ such that $p(\mathbf{x}) = q(L\mathbf{x}, (\mathbf{u}_1, \mathbf{x}), \ldots, (\mathbf{u}_\ell, \mathbf{x}))$.
  Then, for every $1 \leqslant i \leqslant \ell$, using $D_{\mathbf{v}} (\mathbf{u}, \mathbf{x}) = (\mathbf{v}, \mathbf{u})$, we have
  \begin{align*}
    D_{\mathbf{u}_i} p(\mathbf{x}) &= D_{\mathbf{u}_i} q(L\mathbf{x}, (\mathbf{u}_1, \mathbf{x}), \ldots, (\mathbf{u}_\ell, \mathbf{x}))
    = (\mathbf{u}_i, \mathbf{u}_i) \frac{\partial q}{\partial y_{s + i}}(L\mathbf{x}, (\mathbf{u}_1, \mathbf{x}), \ldots, (\mathbf{u}_\ell, \mathbf{x}))\\ &= \frac{\partial q}{\partial y_{s + i}}(L\mathbf{x}, (\mathbf{u}_1, \mathbf{x}), \ldots, (\mathbf{u}_\ell, \mathbf{x})).
  \end{align*}
  Therefore, $q$ does not involve $y_{s + i}$, so we get a representation of $p$ as a polynomial in $L\mathbf{x}$.
\end{proof}

\begin{proposition}\label{prop:jacobian}
  A matrix $L \in \Mat_{s, n}(\mathbb{R})$ is a lumping for a $n$-dimensional system $\dot{\mathbf{x}} = \mathbf{f}(\mathbf{x})$ if and only if, $\forall\mathbf{x} \in \mathbb{R}^n$, 
  $\rspan_{\mathbb{R}}(L)$ is invariant under $J(\mathbf{x})$, the Jacobian matrix of $\mathbf{f}$.
\end{proposition}

\begin{proof}
  We will use the notation from Lemma~\ref{lem:poly_secret}.
  For $\mathbf{v} \in V$,
  \[
  D_{\mathbf{v}} L\mathbf{f}(\mathbf{x}) = \left(\mathbf{v}, \left(\tfrac{\partial}{\partial x_1}, \ldots, \tfrac{\partial}{\partial x_n} \right) \right) L\mathbf{f}(\mathbf{x}) = (LJ(\mathbf{x})) \mathbf{v}.
  \]
  Therefore, Lemma~\ref{lem:poly_secret} implies that $L$ is a lumping of $\dot{\mathbf{x}} = \mathbf{f}(\mathbf{x})$ if and only if $\rspan_\mathbb{R}(LJ(\mathbf{x}))$ is orthogonal to $V$ for every $\mathbf{x}$.
  The latter is equivalent to the invariance of $\rspan_{\mathbb{R}}(L)$ under $J(\mathbf{x})$ for every $\mathbf{x} \in \mathbb{R}^n$.
\end{proof}

%%%%%%%%%%%%%%%%%%%

{
\section{Complexity analysis}\label{sec:complexityanalysis}

\setcounter{lemma}{0}
\renewcommand{\thelemma}{III.\arabic{lemma}}
\setcounter{proposition}{0}
\renewcommand{\theproposition}{III.\arabic{proposition}}
\setcounter{notation}{0}
\renewcommand{\thenotation}{III.\arabic{notation}}
\setcounter{remark}{0}
\renewcommand{\theremark}{III.\arabic{remark}}

In this section, we give upper bounds on the arithmetic complexity (that is, each operation with rational numbers is assumed to have unit cost) of Algorithms~\ref{alg:simplified} and~\ref{alg:inv_subspace} (Propositions~\ref{prop:arithm_comp_inv_subspace} and~\ref{prop:total_complexity}) and their comparison with the complexity bound of the algorithm implemented in ERODE from~\cite[Supporting Information, Theorem~3]{pnas17} (Remark~\ref{rem:comparison_complexity}).

As we explain in Section~IV, our implementation runs Algorithm~\ref{alg:inv_subspace} first and switches to Algorithm~\ref{alg:modular} only if it encounters large numbers (more than $10000$ digits).
For the majority of the models, the switch did not happen.
In these cases, the numbers occurring during the computation will have lengths bounded by the constants, so the arithmetic complexity will be the same as the bit-size complexity, and, therefore, can be used to reason about the runtime.

\begin{remark}\label{rem:sparse}
Before estimating the complexity of the algorithms, we explain the data structures we use for representing vectors and matrices.
\begin{itemize}
    \item \emph{Vectors.} Each vector is represented by an ordered list of indices of the coordinates with nonzero values and by a hashtable with keys being these indices and the values being the values of the corresponding coordinates.
    
    For example, the vector $\mathbf{v} = (0, 0, 3, 1, 0, 0, 5, 1, 0)$ will be represented by the list $(3, 4, 7, 8)$ and hashtable $\{3 \to 3, 4 \to 1, 7 \to 5, 8 \to 1\}$.
    
    If two vectors $\mathbf{v}_1$ and $\mathbf{v}_2$ have $n_1$ and $n_2$ nonzero coordinates, respectively, then their sum and inner product can be computed with expected arithmetic complexitites $\mathcal{O}(n_1 + n_2)$ and $\mathcal{O}(\min(n_1, n_2))$, respectively.
    
    \item \emph{Matrices.} Each matrix is represented as a sparse vector (as described above) of its rows represented also as sparse vectors.
    Then if a matrix $M$ has $n$ nonzero entries, then the product $M \mathbf{v}$ with a sparse vector $\mathbf{v}$ can be computed with expected arithmetic complexity $\mathcal{O}(n)$ by computing inner products of $\mathbf{v}$ with the nonzero rows of $M$,
\end{itemize}
\end{remark}

\begin{proposition}\label{prop:arithm_comp_inv_subspace}
  Let $A$ be a full row rank $s \times n$ matrix over a computable field $\mathbb{K}$ and $M_1, \ldots, M_\ell$ be $n\times n$ nonzero sparse matrices (represented as in Remark~\ref{rem:sparse}) with the total number of nonzero entries being $T$.
  Then the expected arithmetic complexity of Algorithm~\ref{alg:inv_subspace} is $\mathcal{O}(rn(T + r))$ (this is bounded by $\mathcal{O}(n^2 (T + n))$ since $r\leqslant n$), where $r$ is the number of rows in the output.
\end{proposition}

\begin{proof}
  We will analyze the complexity step-by-step.
  The complexity of~\ref{alg2:reduced_row} and~\ref{alg2:pivots} is equal to the complexity of Gaussian elimination, so it can be bounded by $\mathcal{O}(s^2 n)$ field operations (similarly to~\cite[p.~165]{NumLinAlg}).
  \ref{step:while} involves three different operations: computing matrix-vector products, reducing a vector with respect to the rows of $L$, and reducing rows of $L$ with respect to a newly added vector.
  
  We will bound the complexities of these steps separately:
  \begin{itemize}
      \item \emph{Matrix-vector multiplications.} The number of vectors $v$ considered in this step does not exceed the number of pivots in the resulting matrix $L$, which is $r$.
      For each such vector, we multiply it by the matrices $M_1, \ldots, M_\ell$.
      Remark~\ref{rem:sparse} implies that this can be done in $\mathcal{O}(T)$ operations.
      Thus, the total complexity will be $\mathcal{O}(rT)$.
      
      \item \emph{Reducing with respect to the rows of $L$.}
      Consider the vector $v$ from~\ref{step:while}.
      The total number of nonzero entries in $v M_1, \ldots, vM_\ell$ does not exceed $T$.
      Since $L$ is in row reduced echelon form, the total number of elementary row operations used while reducing these vectors with respect to the rows of $L$ will not exceed $T$.
      Each such row operation has complexity $\mathcal{O}(n)$, so the total complexity for the fixed vector $v$ is $\mathcal{O}(nT)$.
      Since there will be at most $r$ such vectors, the overall complexity is $\mathcal{O}(rnT)$.
      
      \item \emph{Reducing rows of $L$ with respect to a newly added vector.}
      There will be $r - s$ newly added vectors. The total number of elementary row operations will be $s + (s + 1) + \ldots + r$. 
      Hence, the total complexity will be $\mathcal{O}((r^2 - s^2)n)$.
  \end{itemize}
  
  Summing up, we obtain
  \[
  \mathcal{O}(s^2n) + \mathcal{O}(rT) + \mathcal{O}(rnT) + \mathcal{O}((r^2 - s^2)n) = \mathcal{O}(rn(T + r)).\qedhere
  \]
\end{proof}

\begin{proposition}\label{prop:total_complexity}
  Consider a system $\dot{\mathbf{x}} = \mathbf{f}(\mathbf{x})$ of $n$ ODEs with polynomial right-hand side.
  Let 
  \begin{itemize}
      \item $M$ be the total number of monomials in the right-hand side;
      \item $p$ be the maximal number of different variables occuring in a monomial;
      \item $r$ be the dimension of the reduced system (so $r \leqslant n$).
  \end{itemize}
  Then the expected arithmetic complexity of Algorithm~\ref{alg:simplified} with~\ref{step:main_alg4} performed by Algorithm~\ref{alg:inv_subspace} is $\mathcal{O}(rn(pM + r))$.
\end{proposition}

\begin{remark}
  If the ODE system represents a chemical reaction network with mass-action kinetics, then $n$ will be the number of species, $M$ will be the number of reactions, and $p$ will be the maximal number of different species among the reactants or products of a reaction.
\end{remark}

\begin{proof}[Proof of Proposition~\ref{prop:total_complexity}]
  We will analyze the complexity of~\ref{step:main_alg1} and~\ref{step:main_alg2} together.
  Each monomial in $\mathbf{f}(\mathbf{x})$ will yield at most $p$ nonzero entries in the matrices $J_1, \ldots, J_N$.
  Therefore, the complexity of constructing these matrices will be $\mathcal{O}(pM)$, and the total number of nonzero entries in these matrices will not exceed $pM$.
  The complexity of~\ref{step:main_alg3} is $\mathcal{O}(1)$.
  Now we apply Proposition~\ref{prop:arithm_comp_inv_subspace} to matrices $J_1, \ldots, J_N$, and obtain that the complexity of~\ref{step:main_alg4} is  $\mathcal{O}(rn(pM + r))$.
  The overall complexity will be  $\mathcal{O}(rn(pM + r))$.
\end{proof}

\begin{remark}[Comparison with ERODE]\label{rem:comparison_complexity}
    The complexity of the algorithm implemented in ERODE given by~\cite[Supporting infomration, Theorem~3]{pnas17} can be written in the notation of Proposition~\ref{prop:total_complexity} as $\mathcal{O}(Mp^2 \ell \log n) \leqslant \mathcal{O}(M^2 p^3 \log n)$ {(this is the worst-case complexity of a deterministic algorithm, so it is also the expected complexity)}, where $\ell$ is the number of distinct partial derivatives among the monomials with different signs (we do not use this parameter in our complexity analysis).
    
    Bringing our bound and this bound to a common set of parameters, we get $\mathcal{O}(n^2(pM + n))$ and $\mathcal{O}(M^2 p^3 \log n)$, respectively.
    These bounds indicate that one algorithm can outperform the other one depending on the parameters of the model considered.
\end{remark}
}

%%%%%%%%%%%%%%%%%%%

\section{Modular decomposition for cartilage breakdown model in~\cite{proctor2014computer}}\label{sec:cartilage}

\begin{figure}[t]
\centering
\includegraphics[width=.9\linewidth]{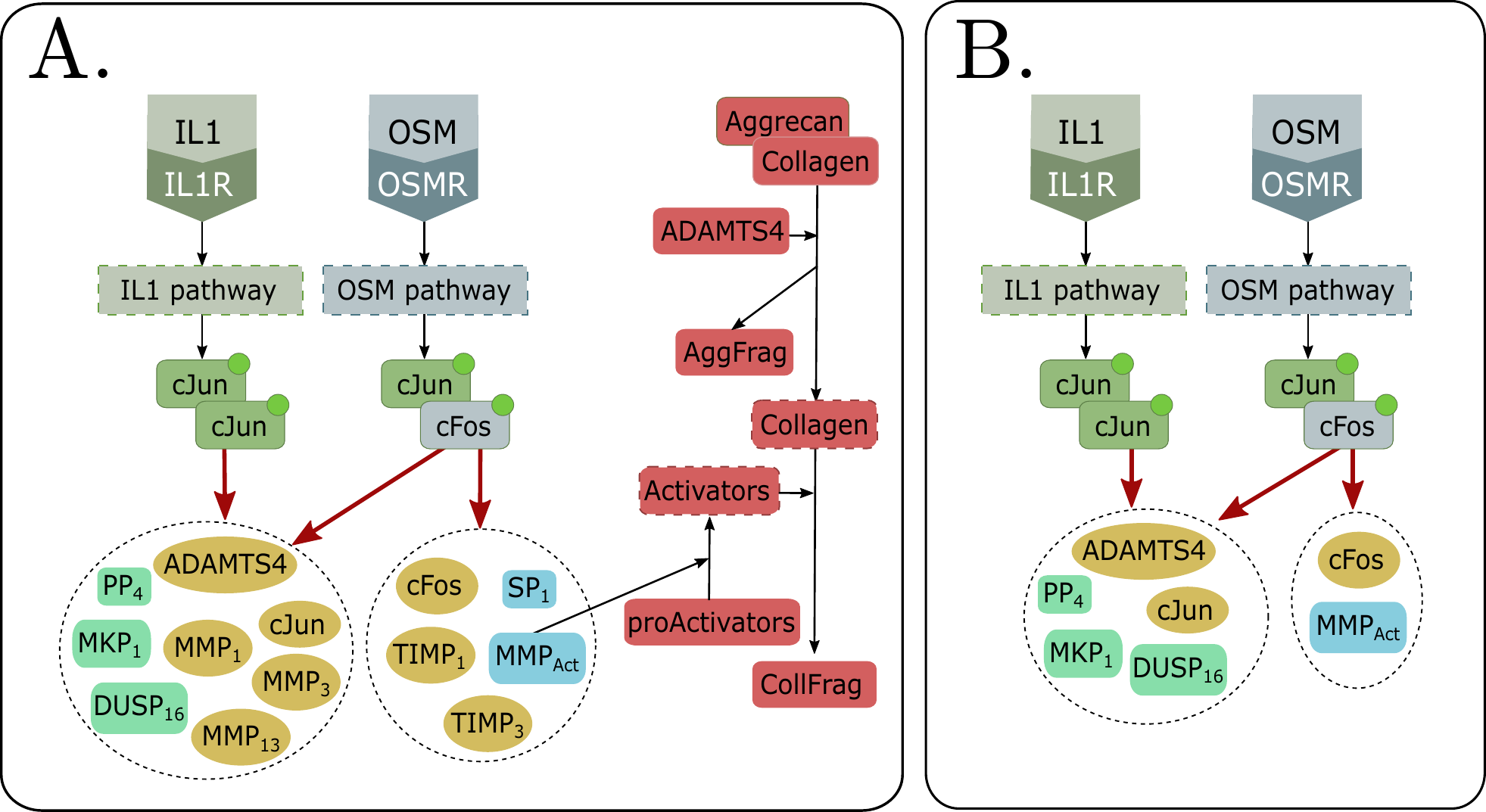}
\caption{
(A) Adaptation of the three molecular pathways from~\cite{proctor2014computer}. (B)~Reduced model obtained while preserving the  phosphorylated forms of \ce{cJun} and \ce{cFos}. Dotted boxes represent abstractions of groups of biochemical species which are not fully shown here to reduce clutter.}
\label{fig:coll}
\end{figure}
This section discusses a pattern of modular decomposition similar to Section 4.1 in the main text, on a model of  cartilage breakdown pathway from~\citep{proctor2014computer}, illustrated in Fig.~\ref{fig:coll}(A). The model is available in the BioModels repository as~\biomodel{BIOMD0000000504}. The system comprises three modules: an Interleukin-1 (\ce{IL1}) signaling pathway, an \ce{OSM} signaling pathway, and a circuit of activation of \ce{proMMPs} that concludes with the degradation of \ce{Aggrecan} and \ce{Collagen}. 

In the first module, \ce{IL1} binds its receptor (\ce{ISMR}) to start a cascade of phosphorylation events (not shown) that activates \ce{cJun}. After dimerization, \ce{cJun} upregulates collagenases \ce{MMP_{\{1,3,13\}}} and phosphatases MKP1, \ce{PP 4}4 and DUSP16. In the second module, \ce{OSM} binds to the receptor \ce{OSMR}; the pathway concludes with the phosphorylation of \ce{cFos}. The active \ce{cFos} can reversibly bind to phosphorylated \ce{cJun} in a complex cJun-cFos which acts as transcription factor and upregulates the transcription factor \ce{SP 1}, \ce{TIMPs_{\{1,3\}}}, cFos, cJun, a generic \ce{MMP_{Activator}} and all the upregulated components from \ce{IL 1} module. In the third module, the Aggrecan-Collagen complex separates due to the interaction with \ce{ADAMTS 4}, and the units of \ce{Aggrecan} in the complex transform into fragments (\ce{AggFrag}). The units of \ce{Collagen} interact with several \ce{Activators} (collagenases such as \ce{MMP_{\{1,3,13\}}} or \ce{MMP_{Act}}) that destroy the protein structure, producing collagen fragments (\ce{CollFrag}).

The original model consists of 74 variables. By
preserving the phosphorylated molecules of \ce{cFos} and \ce{cJun}, which are some of the species of interest in the study by~\citet{proctor2014computer}, CLUE removes the pathway for the decomposition of the Aggrecan-Collagen complex, together with the mRNA variants of \ce{MMP_{\{1,3,13\}}}, \ce{TIMP_{\{1,3\}}}, and \ce{SP1}. The reduced model with 43 variables can be interpreted as the network in Fig.~\ref{fig:coll}~(B). Again, CLUE
simplifies  branches of the pathway that do not affect the dynamics of the observables. The reduction by forward equivalence, instead, collapses only the variables corresponding to the species \ce{Aggrecan}, \ce{AggFrag}, \ce{Collagen}, and \ce{CollFrag}, providing a model with 71 variables. Differently from the previous example, this block collapses end species (\ce{AggFrag} and \ce{CollFrag}) together with an input species (\ce{Aggrecan}) which is assumed to have no dynamics (i.e., zero derivative), as well as a species (\ce{Collagen}) that undergoes degradation.

\section{Comparison of Algorithm~\ref{alg:inv_subspace} and Algorithm~\ref{alg:modular}}\label{sec:runtimes}

\setcounter{lemma}{0}
\renewcommand{\thelemma}{IV.\arabic{lemma}}
\setcounter{proposition}{0}
\renewcommand{\theproposition}{IV.\arabic{proposition}}
\setcounter{notation}{0}
\renewcommand{\thenotation}{IV.\arabic{notation}}
\setcounter{remark}{0}
\renewcommand{\theremark}{IV.\arabic{remark}}

{
As  mentioned in the main text, Algorithm~\ref{alg:inv_subspace} is typically faster for simpler cases, while the performance of Algorithm~\ref{alg:modular} is more robust.
The ratios of the runtime of Algorithm~\ref{alg:modular} and the runtime of Algorithm~\ref{alg:inv_subspace} for an extended set of benchmarks are collected in Table~\ref{tab:Alg2vsAlg3} below.
The value $< 0.01$ refers to the fact that Algorithm~\ref{alg:inv_subspace} has been running for 100 times more than the runtime of Algorithm~\ref{alg:modular} but did not produce any result and has been stopped.
The benchmarks are available in the repository~\url{https://github.com/pogudingleb/CLUE/tree/master/examples}.
For three of the models, we had several sets of observables, the indexes of the sets (as listed in the repository) are given in the parenthesis.

From the table, one can see that Algorithm~\ref{alg:inv_subspace} is faster than Algorithm~\ref{alg:modular} by about a factor of $6$ for the majority of given examples.
Typically, this happens if the dimension of the reduced model is relatively small or the form of reduction is relatively simple.
On the other hand, in the cases in which Algorithm~\ref{alg:inv_subspace} encounters very long integers during the computation (like~\citep{Barua} and~\citep{Faeder01042003} models), it is likely to get stuck while Algorithm~\ref{alg:modular} terminates in reasonable yielding to more than 100-fold speed up.
  
\begin{table}[H]
    \centering
    \begin{tabular}{|l|c|}
    \hline
        Model &  time(Alg.~\ref{alg:modular}) / time(Alg.~\ref{alg:inv_subspace}) \\
    \hline
        \cite{li2006stochastic} & $2.5$ \\ 
    \hline
         \cite{proctor2014computer} (1) & $3.0$\\
    \hline
    \cite{proctor2014computer} (2) & $4.0$\\
    \hline
    \cite{proctor2014computer} (3) & $3.2$\\
    \hline
    \cite{proctor2014computer} (4) & $4.0$\\
    \hline
        \cite{borisov2008domain} & $6.0$ \\
    \hline
        \cite{sneddon2011efficient}, $m = 2$ & $5.0$ \\
    \hline
        \cite{sneddon2011efficient}, $m = 3$ & $5.0$ \\
    \hline
        \cite{sneddon2011efficient}, $m = 4$ & $6.0$ \\
    \hline
        \cite{sneddon2011efficient}, $m = 5$ & $6.7$ \\
    \hline
        \cite{sneddon2011efficient}, $m = 6$ & $6.9$ \\
    \hline
        \cite{sneddon2011efficient}, $m = 7$ & $6.7$ \\
    \hline
        \cite{sneddon2011efficient}, $m = 8$ & $6.6$ \\
    \hline
         \cite{Barua} (1) & $\mathbf{< 0.01}$ \\
    \hline
         \cite{Barua} (1) & $\mathbf{< 0.01}$ \\
    \hline
         \cite{Pepke2010} & $4.0$ \\
    \hline
         \cite{Faeder01042003} (1) & $5.2$ \\
    \hline
         \cite{Faeder01042003} (2) & $\mathbf{< 0.01}$ \\
    \hline
    \cite{Faeder01042003} (3) & $5.8$ \\
    \hline
    \cite{Faeder01042003} (4) & $5.6$ \\
    \hline
    \cite{Faeder01042003} (5) & $6.6$ \\
    \hline
    \end{tabular}
    \caption{The ratio of the runtimes of Algorithm~\ref{alg:modular} and Algorithm~\ref{alg:inv_subspace} for an extended set of benchmarks\\
    The numbers in parenthesis after a reference refer to the index of the chosen set of observables.}
    \label{tab:Alg2vsAlg3}
\end{table}

In our implementation, these algorithms are combined to benefit from their strengths as follows. 
We first run Algorithm~\ref{alg:inv_subspace}, and if the algorithm encounters very long rational numbers (we use $10000$ digits as the threshold), then we stop it
and run Algorithm~\ref{alg:modular} instead.
In the most frequent case of not so long rational numbers, the runtime is the same as that of Algorithm~\ref{alg:inv_subspace}.
In the cases in which using Algorithm~\ref{alg:modular} is preferable, first trying Algorithm~\ref{alg:inv_subspace} in our implementation adds only a small overhead (less than 10\%) compared to running Algorithm~\ref{alg:modular} by itself.

}

\bibliographystyle{abbrvnat}
\bibliography{bibdata}